\documentclass[superscriptaddress,footnoteinbib,showpacs,twocolumn,amssymb,aps]{revtex4-1}

\usepackage{graphicx,dcolumn,bm,amsmath,color,bm}
\usepackage[normalem]{ulem}

\newcommand{\eq}[1]{Eq.~(\ref{#1})}

\newcommand{\eeq}{ \end{equation} }
\newcommand{\beq}{ \begin{equation} }

\newcommand{\bff}{ {\bf f }}

\newcommand{\bu}{ {\bf u }}

\newcommand{\bhu}{ {\bf \hat{u}} }

\newcommand{\bfr}{ {\bf r} }

\newcommand{\bfq}{ {\bf q} }
\newcommand{\bfnp}{ {\bf n}_{\perp} }
\newcommand{\bfqp}{ {\bf q}_{\perp} }
\newcommand{\bfrp}{ {\bf r}_{\perp} }
\newcommand{\bzh}{ {\bf \hat{z}} }
\newcommand{\bn}{ {\bf \hat{n}} }
\newcommand{\bxh}{ {\bf \hat{x}} }
\newcommand{\byh}{ {\bf \hat{y}} }

\newcommand{\kbt}{k_{\rm B}T}

\begin{document}

\title{Following fluctuating signs: anomalous active superdiffusion of swimmers in anisotropic media}
\author{John Toner}
\email{jjt@oregon.edu}
\affiliation{Department of Physics and Institute of Theoretical Science, University of Oregon, Eugene, OR 97403, USA}
\author{Hartmut L\"{o}wen}
\affiliation{Institut f\"{u}r Theoretische Physik II: Weiche Materie,
Heinrich-Heine-Universit\"{a}t D\"{u}sseldorf, Universit\"{a}tsstra{\ss}e 1, 40225 D\"{u}sseldorf, Germany}
\author{Henricus H.  Wensink}
\affiliation{Laboratoire de Physique des Solides - UMR 8502, Universit\'e Paris-Sud  \& CNRS,  91405 Orsay, France}

\date{\today}

\begin{abstract}
 Active  (i.e., self-propelled or swimming) particles moving through an isotropic fluid exhibit conventional diffusive behavior. 
We report anomalous diffusion of an active  particle moving in an anisotropic, nematic background. 
Whilst the translational motion parallel to the nematic director shows ballistic behavior, the long-time 
transverse 
motion is super-diffusive, with an anomalous scaling $\propto t \ln t$ of the mean squared displacement
with time $t$. This behavior is predicted by an 
 analytical theory that we present here,  and is corroborated by 
numerical simulation of active particle diffusion in a simple lattice model for a nematic liquid crystal. 
It is universal for any collection of self-propelled elements (e.g., bacteria or active rods) 
moving in a  nematic background, provided only that the swimmers are sufficiently dilute that their interactions with each other can be neglected, and that they do not perform ``hairpin" turns. 

\end{abstract}

\maketitle

\section{Introduction}

The physics of microswimmers is a flourishing research field,  which has enriched our understanding of non-equilibrium emergent phenomena, and could lead to 
many applications, such as 
controlled particle separation and self-assembly. By and large, most  artificial microswimmers
considered so far are embedded in a simple Newtonian fluid at low Reynolds number \cite{Marcetti_RMP,romanczuk2012,Gompper_Winkler_review,bechinger_volpe2016,hagan_baskaran2016}.
Many microorganisms in their natural environment, however,  are exposed to much more complex media, which are more appropriately described by complex non-Newtonian fluids \cite{Peer_Fischer,MartinezPNAS2014}.
Examples range from  the motion of cilia and spermatozoa
in mucus \cite{ref4,ref7}  to bacteria in the host
tissue \cite{ref6} and nematodes
migrating though soil \cite{ref5}. Recent efforts aimed at gaining a better understanding of the role of the complex environment involve 
studying microswimming in non-Newtonian solvents such as
viscoelastic fluids
\cite{FuPowers2007,Arratia1,Arratia2,Liu,Poon_PNAS,Peer_Fischer,Lauga0,Lauga1,Lauga2,bechinger2016}, in
liquid crystalline environments
\cite{PNAS_Aranson,Lavrentovich2015,Lauga2,Mushenheim,Sagues,Krieger1,Krieger2,trivedi2015}, in the presence of random \cite{peruaniPRL2013} or patterned \cite{volpeSM2011}  obstacles, or in crystalline \cite{Krieger3,Filion,Poon2016} media.
 
Motivated by recent experiments on ``living liquid crystals" (i.e., bacteria swimming in a nematic background)
\cite{PNAS_Aranson,Sagues,PNAS_Aranson,hernandez-sagues2015},
we study here a swimmer in a nematic liquid crystalline solvent,  when the swimming direction is coupled to the local nematic director sufficiently strongly to prevent hairpin turns by the swimmer. 
The nematic background is anisotropic, with a macroscopic nematic director; allowing  the swimmer
motion  to be decomposed into components
parallel and perpendicular to the nematic director. Our interest is in determining the effects of thermal nematic director fluctuations on the swimmer's motion.

We develop a  hydrodynamic theory that describes the universal behavior of a swimmer
with only a short term memory, and {\it any} coupling to the nematic director that tends to locally align its motion along that director. The only limitation of the theory is that it excludes hairpin turns; since these will be very rare for any appreciable coupling of the swimmer velocity to the director, this is not a serious limitation of the theory. 

Our theory predicts that the mean displacement $\Delta r_{\parallel}$ parallel to the nematic director shows ballistic behavior, that is,  the mean parallel displacement behaves as
\beq
\langle \Delta r_\parallel(t) \rangle \propto t \quad ,
\label{ballistic}
\eeq
while the long-time transverse motion is super-diffusive,  with an anomalous scaling 
\beq
\langle |{\bf \Delta r}_{_\perp}(t)|^2 \rangle \propto t \ln t \quad ,
\label{anomdif}
\eeq
of the mean squared displacement with time $t$. These predictions are corroborated by numerical
simulation of a model in which   the nematic background is represented by  ``Lebwohl-Lasher" spins  on a lattice, to which the swimmer is {\it not} confined. 

This superdiffusive behavior is the signature of a new universality class of active diffusion, and provides a dramatic demonstration of how radically the behavior of an active system can differ from its equilibrium counterparts: even adding a {\it single} active element to an otherwise entirely {\it equilibrium} system completely change the scaling of diffusion. Note further that this change of scaling is an {\it inevitable} consequence of the activity; furthermore, the new scaling that results is {\it universal}: it will occur for {\it any}
swimmer in {\it any} otherwise equilibrium nematic, provided only that our very general and plausible assumption of spatiotemporal locality is met.

\section{Hydrodynamic theory}

We will consider a  self-propelled swimmer moving through an otherwise equilibrium, ordered  uniaxial nematic. This swimmer has no memory, or, at best, only a short term memory,  of its past direction of motion. Furthermore, the dynamics of the entire system (nematic plus swimmer) are rotation invariant: that is, the swimmer  carries no internal ``compass"; any preference it exhibits for one direction of motion over any other must arise from the {\it local} nematic director $\bn(\bfr_s(t))$ at the current location $\bfr_s(t)$ of the swimmer. This requirement of locality arises from the physically reasonable assumption that the interactions of the swimmer with the surrounding nematic are short-ranged in space. 
 
The average value of the  instantaneous  velocity $ d \bfr_{s}(t) / dt$ of such a swimmer {\it must} be along $\bn(\bfr_s(t))$; rotation invariance plus locality allow no other direction (except $-\bn(\bfr_s(t))$; we will discuss this option below). Hence, the instantaneous velocity $ d \bfr_{s}(t) / dt $ must be given by
\beq
{d \bfr_{s}(t)\over dt}=v_s\bn(\bfr_{s}(t),t)+\bff(t) \,,
\label{align}
\eeq
where $\bff(t)$ is a zero mean random fluctuation in the  velocity, and $v_s$ is the mean speed of the swimmer. Note that in general $v_s\ne v^0_s$, where  $v^0_s$ is the ``bare", or instantaneous, speed of the swimmer, due to the effects of fluctuations. Indeed, in general, we expect $v_s < v^0_s$. In practice, $v_s$ can only be determined by measuring the mean motion of the swimmer over long times; this will be discussed in more detail below.

The statistics of the fluctuations $\bff$ are also almost completely determined by the requirements of rotation invariance and locality in space and time. In a ``coarse-grained" theory, in which we imagine having averaged our dynamics over time scales long compared to the time of individual molecular ``kicks" experienced by the swimmer, but short compared to the time scales we wish to investigate, $\bff$ can be thought of as a sum of a large number of random molecular kicks at different microscopic times, which are therefore statistically independent. The central limit theorem then tells us that the statistics of $\bff$ should be Gaussian. Its statistics are then completely specified by its two point correlations with the local nematic director $\bn(\bfr,t)$ and itself; rotation invariance and spatio-temporal locality  imply
that these are given by:
\begin{eqnarray}
  \langle f_{\alpha}(t)f_{\beta}(t')\rangle &= 2\Delta_{I}
\delta_{\alpha\beta}\delta(t-t') +2\Delta_{A} n_{\alpha}(\bfr_{s}(t),t) \nonumber \\ 
&\times  n_{\beta}(\bfr_{s}(t),t)   \delta(t-t') \,,
\label{ffcorr}
\end{eqnarray}
and
\begin{eqnarray}
   \langle f_{\alpha}(t)n_{\beta}(\bfr_{s}(t'),t')\rangle=2\Delta_{fn}
\delta_{\alpha\beta}\delta(t-t')\,,
\label{fncorr}
\end{eqnarray}
where $\alpha$ and $\beta$ are Cartesian indices, and $\Delta_{I}$, $\Delta_{A}$, and $\Delta_{fn}$ are phenomenological parameters which set the size of the fluctuations of the swimmer. Because the swimmer is a non-equilibrium agent, these parameters do not, in general, satisfy any kind of fluctuation-dissipation theorem; they are independent parameters. 

To complete our description, we need to specify the  dynamics of $\bn$. In our simulations, we discretize space into a simple cubic lattice, with sites labeled by $i$,  and take the dynamics of the director far from the swimmer to be purely  relaxational and equilibrium; that is,   
\beq
{d \bn_{i} (t)\over dt} =-\Gamma {\partial H\over\partial \bn_{i}}+\boldsymbol{\zeta}_i(t)\,,
\label{EOMlattice}
\eeq
where the Hamiltonian $H$ is the  discrete Lebwohl-Lasher model \cite{lebwohl} on a simple cubic lattice of lattice constant $a$
\beq
H = -\varepsilon  \sum_{<ij>}  {\mathcal P}_{2} ( \bn_{i} \cdot \bn_{j} ) 
\label{LL}
\eeq
where ${\mathcal P}(x) = (3 x^2 -1)/2$ is the second Legendre polynomial and $\varepsilon$ is a coupling parameter setting the strength of the aligning interactions (considering nearest neighbors only). 

In the continuum limit, this equation of motion for the director becomes
\beq
{\partial \bn(\bfr,t)\over \partial t} =-\Gamma {\delta H_F\over\delta \bn}+\boldsymbol{\zeta}(\bfr,t)\,,
\label{EOMcont}
\eeq
where the continuum Hamiltonian for an equilibrium nematic is, in general,  the well-known Frank free energy \cite{deGennes}
\begin{eqnarray}
H_F &=& \frac{1}{2}\int d^{3} r [K_{1} \left(\nabla\cdot \bn \right)^2+ K_{2}\left( \bn \cdot \left( \nabla \times \bn \right)\right)^2 \nonumber \\
&& \>\>\>\>\>\>\>\>\>\>\>\>\>\>\>\> + K_{3}\left| \bn \times \left( \nabla \times \bn \right)\right|^2  + \lambda(\bfr) |\bn|^{2}]\,,
\label{Frank FE}
\end{eqnarray}
where $K_{1,2,3}$ are the splay, twist, and bend Frank elastic constants, respectively, and $\lambda(\bfr)$ is a Lagrange multiplier that enforces the constraint $|\bn | =1$. For the special case of the Lebwohl-Lasher model, all three Frank constants are equal: $K_1=K_2=K_3\equiv K(T)$, and, as temperature $T\rightarrow 0$, $K(T\rightarrow0)\rightarrow 3\varepsilon/ a$ \cite{priest1972}.

Since the nematic itself, in the absence of the swimmer, is an equilibrium system, the noise $\zeta$ in \eq{EOMcont} must obey the fluctuation-dissipation theorem, which implies: 
\begin{eqnarray}
   \langle \zeta_{\alpha}(\bfr,t)\zeta_{\beta}(\bfr',t')\rangle=2\Gamma\kbt
\delta_{\alpha\beta}\delta^3(\bfr-\bfr')\delta(t-t') \,.
\label{zetacorr}
\end{eqnarray}
The dynamics we have just described are a simplification of those of real bulk nematics \cite{Forster}; those real dynamics are complicated by the coupling of the nematic director to background fluid flow. Nonetheless, the two features of the dynamics which are essential to our calculations, namely i) that equal time correlations are given by the Boltzmann weight associated with the Frank free energy \eq{Frank FE}, and ii) that the dynamics are purely diffusive,  persist in real nematics. The only difference is that in real nematics, there are  {\it two} coupled shear flow-nematoelastic diffusive modes, rather than the single mode that occurs in our model. This difference affects none of our results on anomalous diffusion {\it at all}, as will become clear when we analyze our model. Furthermore, the true director dynamics of real nematics simplifies to our model when inertial effects are negligible which they are whenever the dimensionless parameter $\kappa\equiv K / \rho\nu^2 \ll1$, where $\nu$ is a typical  kinematic shear viscosity (of which there are three in a nematic, due to its anisotropy), $K$ the largest of the Frank constants, and $\rho$ the mass density of the nematic.  In this limit, the shear flow mode and director realignment decouple, and the 
director dynamics is described precisely by a slightly anisotropic version of   our model with kinetic coefficient $\Gamma=1 / 
\rho\nu$. For  most experimentally known nematics, $\kappa \lesssim 10^{-4}$ \cite{Forster}, so 
this is an excellent approximation.

We expect  the effect of the swimmer on the nematic to be negligible, because it is purely local, while the long time behavior of the swimmer is, as we will see,  determined by the long distance correlations of $\bn$. In our simulations, we check this by  including  interactions between the swimmer and the director that locally realign the director as well as the swimmer. As expected, we find that, as we just argued, such swimmer induced director realignments do not affect our results, in the sense that the theory presented here, which ignores them, recovers the observed anomalous diffusive behavior of the lateral motion of the swimmer.

Our complete hydrodynamic theory is thus embodied in equations of motion \eq{align} and \eq{EOMcont} for the swimmer and the nematic director, respectively, supplemented by the expressions \eq{ffcorr} and \eq{fncorr} for the two point correlations of the Gaussian random velocity, and by the Frank free energy  \eq{Frank FE} for the energetics of the nematic director.

Thus, the behavior of the swimmer is completely specified by four equilibrium parameters: the temperature $T$, the three Frank constants $K_{1,2,3}$, and the director kinetic coefficient $\Gamma$, as well as four fundamentally non-equilibrium parameters associated with the swimmer: the mean swimming speed $v_s$, and the three  non-equilibrium noise strengths  $\Delta_{I}$,$\Delta_{A}$,$\Delta_{fn}$, and $D$. As we shall see, the anomalous diffusion is determined entirely by one combination of $v_s$, $T$, $K_{1,2}$, and is 
independent of  the bend Frank constant $K_3$ and of all of the non-equilibrium parameters except the swimming speed $v_s$. Since the form of  \eq{align}, \eq{EOMcont}, \eq{ffcorr},  \eq{fncorr}, and  \eq{Frank FE} are dictated by symmetry and spatio-temporal locality, they are completely universal; that is, they describe {\it any} swimmer in {\it any} nematic. This implies in turn that the conclusions we are about to draw from these equations, in particular, that the swimmer exhibits superdiffusive lateral motion, are universal as well.

There is, however, one limitation on our equations: they exclude ``hairpin turns". These are fluctuations in which the swimmer reverses its direction of motion relative to the local nematic director (that is, where it makes an angle of more than $90^o$ with the director. Such turns are important because of the nematic symmetry of the background nematic, whose implications we will now discuss.

 The nematic phase is  {\it apolar}; that is, in it, although the long axes of the molecules align,  their heads and tails do not.  
This means that reversing the nematic director - i.e., taking $\bn\rightarrow -\bn$ - cannot change anything physical. Our fundamental equation of motion \eq{align} obviously violates this symmetry.
There is, however, a natural way to eliminate this arbitrariness of the sign of $\bn$: we can simply choose the sign of $\bn$ at every point to be such that it makes an angle of less than $180^\circ$
with the initial direction of the swimmer's velocity.

However, nematic symmetry also implies our equation of motion \eq{align}  {\it cannot} continue to hold  once the swimmer makes a  ``hairpin" turn: that is, once its direction of motion makes an angle of more than  $>90^\circ$  with the local nematic director $\bn$. Rather, since, in a nematic state, the swimming velocity can only align with the nematic {\it axis}, we would expect that, once a hairpin turn has occurred, the swimmer will now seek to align, not with $\bn$ as defined above, but with $-\bn$. An extension of our model to allow for this effect finds  that all of the behavior we find below will be cut off for $t \gg t_{\text{hairpin}}$, where $t_{\text{hairpin}}$ is the mean time between hairpin turns.  This argument is discussed in detail in  Appendix B.  Fortunately, this time $t_{\text{hairpin}}$ can be made exponentially long: we would expect it to scale like $\exp(\Delta E/\kbt)$, where $\Delta E$ is the height of the ``energy barrier" against a reversal of the swimmer direction of motion (that is, the energy cost of the swimmer making an angle of $90^o$ with the local nematic axis).  This can therefore be made very long in a model simply  by  making $\Delta E\gg\kbt$, as we have done in our simulations which we will specify in the subsequent paragraph.  Indeed, we have {\it never} observed a hairpin turn in our simulations. More importantly, we also expect that, deep within the nematic phase and for a strongly aligned swimmer, $\Delta E\gg\kbt$, so hairpins should be rare, if not non-existent, in many real experiments as well.
Hairpin turns can also be avoided by considering not a self-propelled particle, but a sedimenting one: that is, a particle whose motion is driven by an external force, like gravity, or electric or magnetic fields. This also gives one the option of studying motion that is directed in a different direction than that of nematic alignment. We will discuss this interesting problem in a future publication.


We now proceed to analyze the implications of this theory for the motion of the swimmer.
We'll start with the mean motion.
Taking the average of  \eq{align}, and recalling that $\langle\bff\rangle={\bf 0}$, we immediately obtain an expression for the mean position of the swimmer: 
\begin{eqnarray}
\langle\bfr_{s}(t) \rangle =v_st\langle\bn\rangle\equiv v_zt\bzh\,,
\label{meanrs}
\end{eqnarray}
where 
we have taken the mean direction of the nematic director $\bn$ to be along $\bzh$, and 
 the mean swimmer speed in the $z$ direction is
given by $v_z=v_s|\langle\bn\rangle|$ 
Thus, the mean motion of the swimmer is purely ballistic. The speed $v_z$ of this motion can {\it not} be determined by the continuum theory used here, since fluctuations of the director 
away from $\bzh$, which reduce $\langle\bn\rangle$ below $1$, are, in three dimensions,  dominated by short-wavelength fluctuations, which are not accurately described by the continuum, long-wavelength Frank free energy \eq{Frank FE}. This domination by short wavelengths can be seen by noting that, roughly speaking, the mean squared fluctuations in Fourier space $\langle|\bfnp(\bfq)|^2\rangle$ of the components of the director perpendicular to $\bzh$ predicted by the Boltzmann weight associated with the Frank free energy \eq{Frank FE} obey $\langle|\bfnp(\bfq)|^2\rangle\propto1 /q^2 $. Since  the $\bfq$ space volume  in a spherical shell $q_0\le|\bfq \le 2q_0$ scales like $q_0^3$ in $d=3$, while the typical $\langle|\bfnp(\bfq)|^2\rangle$ in that shell scales like $1 / q_0^2$, the total contribution of such a shell to the mean squared real space fluctuations  $\langle|\bfnp(\bfr)|^2\rangle$, which contribution is proportional to $\int_{q_0\le|\bfq|\le2q_0} d^3 q \,  \left< |\bfnp \left(\bfq, t\right)|^2\right>$,  grows linearly with  
$q_0$. That is, regions of larger $\bfq$ (i.e., smaller wavelength $1/ |\bfq| $) contribute more to $\langle|\bfnp(\bfr)|^2\rangle$ than regions of smaller $\bfq$. Hence, we can not compute these fluctuations from a long wavelength theory. We therefore cannot compute $\langle\bn\rangle$, and, therefore, cannot compute $v_z$. We must instead take it as yet another phenomenological parameter of our model. 
Equivalently, if we incorporate short wavelength effects by introducing an ultraviolet cutoff $\Lambda$ to our wavevector integrals, the value of $\langle\bn\rangle$, and, therefore, of $v_z$, will depend on $\Lambda$, which is another parameter. 

Nonetheless, we have still made a  universal scaling prediction: the mean motion of the swimmer is ballistic, as shown by  \eq{meanrs}.

We now turn to the fluctuations about this mean. Consider first  the mean squared lateral displacement of the swimmer:
\begin{eqnarray}
 \langle ( \Delta \bfr_{s}^{\perp}(t))^{2} \rangle \equiv \left<\left| \bfr_{s}^{\perp}(t) -
 \bfr_{s}^{\perp}(0) \right|^2
\right>
\label{RW1}
\end{eqnarray}
{\it perpendicular} to the mean director of the nematic. Here
and throughout this paper, $\perp$ and $z$ denote directions
perpendicular to, and  along, the nematic director, respectively.

Using the projection of our equation of motion \eq{align} perpendicular to the mean nematic direction $\bzh$, which reads
\beq
{d \bfr_{s}^{\perp}(t)\over dt}=v_s\bfnp(\bfr_{s},t)+\bff_{_\perp} \,,
\label{alignperp}
\eeq
Integrating \eq{alignperp} gives
\beq
\Delta\bfr_{s}^{\perp}(t)\equiv\bfr_{s}^{\perp}(t)-\bfr_{s}^{\perp}(0)=\int_0^tdt'\left(v_s\bfnp(\bfr_{s},t')+\bff_{_\perp}(t')\right) \,.
\label{delrperp}
\eeq
Squaring this, and averaging,  we find that $ \langle ( \Delta \bfr_{s}^{\perp}(t))^{2} \rangle$ is given by
\begin{widetext}
\begin{eqnarray}
 \langle ( \Delta \bfr_{s}^{\perp}(t))^{2} \rangle =\int^t_0 d t^{\prime} \int^t_0 dt^{\prime\prime}
\left[ v_s^2\left<\bfnp(  \bfr_{s}(t^{\prime}),t^{\prime}) \cdot \bfnp(  \bfr_{s}
(t^{\prime\prime}), t^{\prime\prime})\right>+2v_s\left<\bfnp(  \bfr_{s}(t^{\prime}),t^{\prime}) \cdot  \bff_{_\perp}
(t^{\prime\prime})\right>+\left<\bff_{_\perp}
(t') \cdot \bff_{_\perp}
(t^{\prime\prime})\right>\right] \quad .
\label{RW 3}
\end{eqnarray}
\end{widetext}
 Using  the expressions \eq{ffcorr} and \eq{fncorr} for the two-point correlations of the Gaussian random velocity, we can immediately evaluate the last two terms,  denoted by $I_{2}$ and $I_{3}$, respectively. The first of them is
\beq
I_2 =\int^t_0 d t^{\prime} \int^t_0 dt^{\prime\prime}
2v_s\left<\bfnp(  \bfr_{s}(t^{\prime}),t^{\prime}) \cdot  \bff_{_\perp}
(t^{\prime\prime})\right>= 6\Delta_{fn}t\quad,
\label{nfcont1}
\eeq
while the second is
\beq
I_3 =\int^t_0 d t^{\prime} \int^t_0 dt^{\prime\prime}
\left<\bff_{_\perp}
(t') \cdot \bff_{_\perp}
(t^{\prime\prime})\right>= \left[6\Delta_{I}
+2\Delta_{A}
\right]t
\label{nfcont2}
\eeq
Both of these terms are extremely boring: their contribution to 
the mean squared lateral wandering $ \langle ( \Delta \bfr_{s}^{\perp}(t))^{2} \rangle$ is simply conventionally diffusive: that is, proportional to time $t$. The anomalous diffusion that we predict comes entirely from the first term in  \eq{RW 3}:
\begin{eqnarray}
I_1 =v_s^2\int^t_0 d t^{\prime} \int^t_0 dt^{\prime\prime}
\left<\bfnp(  \bfr_{s}(t^{\prime}),t^{\prime}) \cdot \bfnp(  \bfr_{s}
(t^{\prime\prime}), t^{\prime\prime})\right> \,.
\nonumber\\
\label{RW 3.1}
\end{eqnarray}
Because the nematic dynamics are invariant under space and time translations, the general director two point correlation function  depends only on the differences of the space and time coordinates; that is
\begin{eqnarray}
C_\perp\equiv\left<\bfnp(  \bfr',t') \cdot \bfnp(  \bfr'', t^{\prime\prime})\right> = C_{_\perp} \left(\bfr' -
 \bfr'', t- t^{\prime}\right)\nonumber\\
 \label{nemcorr}
\end{eqnarray}
Now in \eq{RW 3.1}, we need this correlation function evaluated when $\bfr'=\bfr_s(t')$ and $\bfr''=\bfr_s(t'')$. These vectors are given by:
\begin{eqnarray}
\bfr_s (t) = \bfr_s (0) + v_st \bzh +
\Delta\bfr_{s}^\perp(t)  \quad  .
\label{RW 5}
\end{eqnarray}
 We will show a posteriori that the typical size of $\bfrp$, as determined by its root mean squared value $\sqrt{\left<|\Delta\bfr_s^\perp(t)|^2\right>}$, is always much less than $v_st$ as $t\rightarrow\infty$. Therefore, since $C_{_\perp} \left(\bfr' -
 \bfr'', t- t^{\prime}\right)$ is a roughly isotropic  function of the relative position 
vector $\bfr^{\prime} - \bfr_{s}^{\prime\prime}$ \cite{deGennes} \footnote{Indeed, in the one Frank constant approximation, which is appropriate for the Lebwohl-Lasher model we simulate, it is perfectly isotropic.}
 we can neglect its $\bfr_s^\perp$ component in \eq{RW 3.1}.  This leads to the simplifying approximation 
\beq
\bfr_s (t') - \bfr_s (t'') \approx v_s(t'-t'')\bzh  \quad  .
\label{drz}
\eeq
Furthermore, because of the slow (diffusive) dynamics of the nematic, we can neglect $t-t'$ as well. To see this, note that  $C_{_\perp} \left(\bfr' -
 \bfr'', t- t^{\prime}\right)$ can only change substantially from its value at $t-t'=0$ when $t-t'$ is large enough for diffusion to occur from $\bfr'$ to $\bfr''$. This requires 
$\sqrt{D|t'-t''|}\gtrsim|\bfr' -
 \bfr''|$,  where  $D$ is  the nematic diffusion constant, which is given for our simple relaxational model (\ref{EOMcont}) by $D = \Gamma K = K/ \rho \nu$. We stress that the anisotropy of the Frank free energy (\ref{Frank FE}) when the Frank constants $K_{1,2,3}$ are unequal leads to anisotropic diffusion; this does not, however, affect the argument we are presenting here, which depends only on the diffusive scaling, and holds even for anisotropic diffusion.
For the Lebwohl-Lasher spin model, this diffusion constant  $D \sim Ka^{3}/\xi \sim 3\varepsilon a^{2} / \xi $, where $\varepsilon$ represents the spin-spin coupling strength and  $\xi$ the spin rotational friction coefficient  (details will be given in the next paragraph), while for real 
nematics, it is given by  $D\sim K / \rho \nu$. Using  \eq{drz} in this condition implies $\sqrt{D|t'-t''|}\gtrsim v_s |t'-t''|$, which is only satisfied for small time differences $|t'-t''|$; specifically, for $|t'-t''|\lesssim t_0$, where $t_0=D/v_s^2$. On longer timescales, i.e., $|t'-t''|\gtrsim t_0$, the correlation function in \eq{RW 3.1} can be replaced by its value at $t'-t''=0$; that
is, we can use the {\it equal-time} correlation function
\begin{eqnarray}
C_\perp(\delta t)=\left \langle \bfnp \left(v_s\delta t \bzh, 0\right)  \cdot \bfnp \left({\bf 0},  0 \right)\right \rangle \quad ,
\label{Cn1}
\end{eqnarray}
where $\delta t\equiv t'-t''$, 
in place of the full nematic corelation in \eq{RW 3.1} for $|t'-t''|\gtrsim t_0$. 

This is very convenient, since equal-time correlations can be calculated for the nematic simply using the Boltzmann weight associated with the Frank free energy \eq{Frank FE}. Expressing $C_\perp$ in terms of its spatio-temporal Fourier transform  \eq{neq}  gives
\beq
\left \langle \bfnp \left(v_s\delta t \bzh, 0\right)  \cdot \bfnp \left({\bf 0}, 0\right)\right \rangle =\int \frac{d^2 q_{_\perp} dq_z}{(2\pi)^3} \, e^{iv_s q_z \delta t} \left \langle |\bfnp \left(\bfq\right)|^2\right \rangle
\label{ncorrft1}
\eeq
where the equal time, equilibrium, spatially Fourier transformed  correlation function $\left< |\bfnp \left(\bfq\right)|^2\right>$  
can easily be evaluated from the Boltzmann weight associated with the Frank free energy, and is \cite{deGennes}:
\begin{eqnarray}
 \left \langle |\bfnp \left(\bfq\right)|^2\right \rangle={k_BT
\over K_1 q_{\perp}^2+K_3q_z^2}
+{k_BT \over K_2 q_{\perp}^2+K_3q_z^2}
\label{neq}
\end{eqnarray}
Performing the integral over $q_z$ by
complex contour  techniques gives
\begin{eqnarray}
C(\delta t) &=& \left \langle \bfnp \left(v_s\delta t \bzh, 0\right)  \cdot \bfnp \left({\bf 0}, 0\right)\right \rangle \nonumber \\ 
&=& k_BT\int \frac{d^2 q_{\perp}}{8\pi^2K_3} \left({e^{-v_s \gamma_1q_{\perp} |\delta t|}\over\gamma_1q_{\perp}}+{e^{-v_s \gamma_2q_{\perp} |\delta t|}\over\gamma_2q_{\perp}}\right) \quad   \label{ncorrft1}
\end{eqnarray}
where  we have defined $\gamma_{{1,2}} \equiv\sqrt{ K_{1,2} / K_3}$. Doing the simple integral over $\bfqp$ then gives:
\beq
C(\delta t) = \frac{ k_{B}T (K_{1}^{-1}+K_{2}^{-1})} { 4\pi v_{s} |\delta t|}\,.
 \label{ncorrrealt}
\eeq
Oddly, this doesn't depend on $K_3$. Even odder, it remains finite and non-zero when $K_{2,3}\rightarrow\infty$, which suggests that the anomaly persists even in a smectic liquid crystal. We have verified that this is true by treating the smectic explicitly; the result is somewhat different from  \eq{ncorrrealt} for subtle reasons that we will discuss later, but the essential phenomenon of anomalous diffusion persists.

Note that this decay of correlations is extremely slow; it is this slow decay, as we will see in a moment, that is responsible for the anomalous diffusion.
What is happening here physically is that, although the swimmer has no long term memory,  the nematic does. This long term nematic memory comes from the fact that the nematic has Goldstone modes, which relax slowly; indeed, they relax at a rate that vanishes as their length scale goes to infinity, which is why they can give rise to such a long $1/\delta t$ tail in their correlations.

It is important to note that this scaling law for $C(\delta t)$ only holds for $\delta t$ large, since it is only for such times that the hydrodynamic theory is valid. Thus, we are not concerned with any apparent divergences at short times 
that occur when 
\eq{ncorrrealt} is inserted into  \eq{RW 3}. Divergences as  $t\rightarrow\infty$ {\it are} real, on the other hand.

Inserting \eq{ncorrrealt}  into  \eq{RW 3} gives 
\begin{widetext}
\begin{eqnarray}
 \langle ( \Delta \bfr_{s}^{\perp}(t))^{2} \rangle = \frac{v_{s} k_{B}T}{4\pi} (K_{1}^{-1}+K_{2}^{-1})\int^t_0 d t^{\prime} \left[\int^{t^\prime-t_0}_0 {dt^{\prime\prime}\over t^\prime-t^{\prime\prime}}+\int^{t^\prime+t_0}_0 {dt^{\prime\prime}\over t^{\prime\prime}-t^\prime}+{\mathcal O}(1)  \right]+D'_0t\quad ,\nonumber\\
\label{RW 6}
\end{eqnarray}
\end{widetext}
where the $D'_0$ term incorporates the ``boring", linear in $t$ contributions of $I_{2,3}$, and the ${\mathcal O}(1)$ represents the contributions from $C(\delta t)$ for $\delta t\lesssim t_0$. The latter clearly also gives rise to another boring contribution to $ \langle ( \Delta \bfr_{s}^{\perp}(t))^{2} \rangle$
proportional to $t$. We will lump that contribution together with the $D'_0t$ term, and call the result $D_0t$. The interesting contribution comes from the explicitly displayed $1/ |t'-t''|$ terms in \eq{RW 6}; evaluating those integrals gives
\beq
 \langle ( \Delta \bfr_{s}^{\perp}(t))^{2} \rangle = \frac{v_{s} k_{B}T}{2\pi} (K_{1}^{-1}+K_{2}^{-1}) t \ln \left(\frac{t}{  t_{0}}\right) + D_0t
 \label{theo}
\eeq
\eq{theo} is the fundamental result of this work. Its form clearly demonstrates how the combination of activity (embodied in the swimming speed $v_s$) and the Goldstone mode fluctuations of an otherwise equilibrium system (manifest in the appearance of $\kbt$ and the Frank constants $K_{1,2}$), leads to a fundamentally different scaling behavior of the 
random motion of the swimmer from that found in {\it any} equilibrium system, since equilibrium systems will always exhibit diffusive behavior  $\langle ( \Delta \bfr_{s}^{\perp}(t))^{2} \rangle=D_{s}t$ behavior, while our $t\ln t$ behavior cannot be written in that form, unless one defines a time-dependent "renormalized diffusion coefficient" $D_{s}\propto \ln t$ which, surprisingly, {\it diverges} as $t\rightarrow\infty$. 

What is particularly striking about our result is that this radically different non-equilibrium scaling is arising due to the addition of a {\it single} non-equilibrium element (the swimmer itself) to an otherwise entirely equilibrium model. Indeed, even the rotational motion of the swimmer is effectively equilibrium rotational diffusion plus equilibrium alignment with the nematic director. It is {\it only} the self-propulsion ($v_s$) that makes the system non-equilibrium; yet this is sufficient to lead to an infinite, non-equilibrium renormalization of the diffusion coefficient.

 \begin{figure} \includegraphics[width=0.9 \columnwidth]{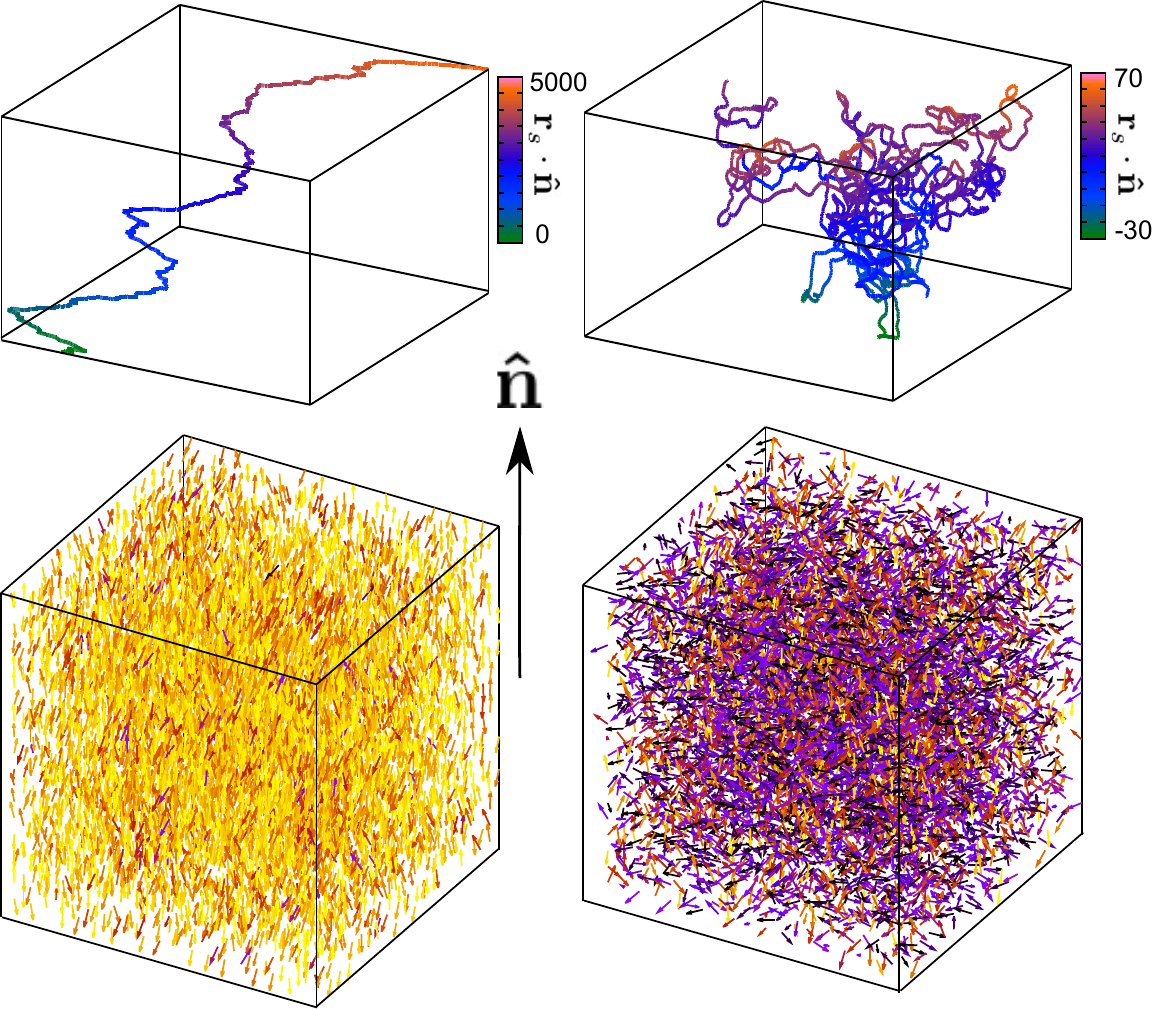}
\caption{Simulation snapshots showing diffusive trajectories of an active particle embedded in a lattice liquid crystal of $N=50^{3}$ Lebwohl-Lasher
spins with color-coded orientations. Both swimmer trajactories cover a time
interval of $\tau=500$. The bottom panels represent a cut-out of the total
simulation box. The boxes enveloping the trajectories are not to scale.
The left trajectory corresponds to persistent active diffusion along the
director of a nematic phase at temperature $T^{\ast} = 0.33$. The swimmer
performs no ``hairpin'' turns. The trajectory on the right displays random
diffusion of an active particle moving in an isotropic phase at $T^{\ast}
= 5$.} 
\label{fig1} 
\end{figure} 

\section{Simulation of an active particle in a nematic liquid crystal}
We will now test the theoretical predictions by numerical simulation. We model the coupled system of an active particle and the nematic background by a continuous overdamped dynamics of the swimmer and a discrete Lebwohl-Lasher lattice model for the nematic background. The bulk properties of the latter are well-known
in the absence of the swimmer involving  an isotropic-nematic (IN) phase transition.

The active particle is characterized
by its center-of-mass position $\bfr_{s}$ and orientation vector $\bhu_{s}$ which describes the swimming direction.
The motion of the swimmer is prescribed by the overdamped Langevin equations:
\begin{align}
d_{t} \bfr_{s}  &= \zeta_{s}^{-1}   F_{0} \bhu_{s}   \nonumber \\ 
d_{t} \bhu_{s} &=  \xi_{s}^{-1}  ( {\bf \Omega}_{s} + {\bf \hat{\Omega}}_{s} ) \times \bhu_{s}  
\end{align}
with $\{ \zeta_{s}, \xi_{s} \} $ (effective) translational and rotational friction factors, $F_{0}$ an effective active force. 
For most swimmers the translation fluctuations exerted by the environment should be of minor importance compared to the orientational noise the objects experience (due to e.g. flagellar motion) \cite{2011DrescherEtAl} and we shall neglect translational noise. 
${\bf \hat{\Omega}}_{s}$ is an intrinsic Gaussian torque
and ${\bf \Omega}_{s}$ an effective torque 
arising from the coupling of the swimmer to the nematic medium.
The nematic background is described
by a cubic lattice of `spins' with positions $\{ \bfr_{i} \} $ and unit vectors $\{ \bhu_{i} \} $
describing the spin orientations.  
The liquid crystalline background is described within the discrete Lebwohl-Lasher  model  \cite{lebwohl, allen2005} on a simple cubic lattice 
with the Hamiltonian given by \eq{LL}. 

We define the effective temperature $T^{\ast} = k_{B}T / \varepsilon $ of the liquid crystalline medium. 
The Lebwohl-Lasher model exhibits a first-order isotropic-nematic transition at an effective temperature 
 $T ^{\ast} = 1.12$ \cite{fabbri1986}.  For the dynamics of the individual classical spins $\{ \bhu_{i} \} $,
 we also assume overdamped Langevin-like dynamics resulting in:
\beq
 d_{t} \bhu_{i} (t) = \xi^{-1} ( {\bf  \Omega}_{i}  +  {\bf \Omega}_{i,s}  + {\bf \hat{\Omega}}_{i} ) \times \bhu_{i}
 \label{EOMsim}
\eeq
with spin rotational friction $\xi$  and ${\bf \hat{\Omega}}_i$ a random torque, for which we take the usual white noise 
characteristics $\langle \hat{\Omega}_{\alpha} \rangle =0 $ and 
$\langle \hat{\Omega}_{\alpha} (t) \hat{\Omega}_{\beta} (t^{\prime}) \rangle = 2 k_{B}T \xi \delta_{\alpha \beta}  \delta(t - t^{\prime})$
to ensure that the system is kept at temperature $T$ in the absence of the swimmer. 
  Here ${\bf \Omega}_{i}\times\bhu_i = \lambda_i\bhu_i-\partial H/\partial \bhu_{i}$ (the Lagrange multiplier $\lambda_i$ enforces the fixed length constraint $|\bhu_i| =1$)  denotes the torque exerted on spin $i$ by its nearest neighbors   (denoted by$\langle j \rangle$):
 \beq
 {\bf \Omega}_{i} =  3 \varepsilon   \sum_{\langle j \rangle} (\bhu_{i} \cdot \bhu_{j}) (\bhu_{i} \times \bhu_{j})
  \label{p2torque}
  \eeq
In addition, the active particle experiences a torque ${\bm \Omega}_{i,s}$ exerted by the surrounding spins. For simplicity we take a ${\mathcal P}_{2}$ coupling with strength $\varepsilon_{s}$ (cf. \eq{p2torque})
 \beq
 {\bf \Omega}_{s} =  3 \varepsilon_{s}   \sum_{j} (\bhu_{s} \cdot \bhu_{j}) (\bhu_{s} \times \bhu_{j}) g({\bf R}_j-\bfr_s)
 \label{swimtorque}
  \eeq
Assuming that  the total system is torque-free, we require that the torque exerted by the particle onto the spins be of equal amplitude but opposite sign, so that  \ ${\bf \Omega}_{i,s}=-{\bf \Omega}_{s}$. The function $g(R_j)$  specifies the dependence  of the swimmer-spin interaction on the distance $R_{j}$ between the spin $j$ and the active particle. The function is taken to be an exponentially damped  one, namely $g(r) = \exp[-(r / \sigma)^{2}]$ with a characteristic decay length $\sigma$. The source of
damping could stem from the non-Newtonian nature of the liquid crystal
solvent  or the
presence of no-slip boundaries. The decay length $\sigma$ is typically larger than the lattice constant $a$ 
such that discretization effects can be ignored.   Fixing $\varepsilon_{s} = \varepsilon$ guarantees a strong orientational coupling between the swimmer and the director to the extent that  
tumbling events, characterized by the swimmer orientation making a hairpin turn, are extremely rare. This point is made explicit in Appendix A.

 \begin{figure*}
\includegraphics[width=1.8 \columnwidth]{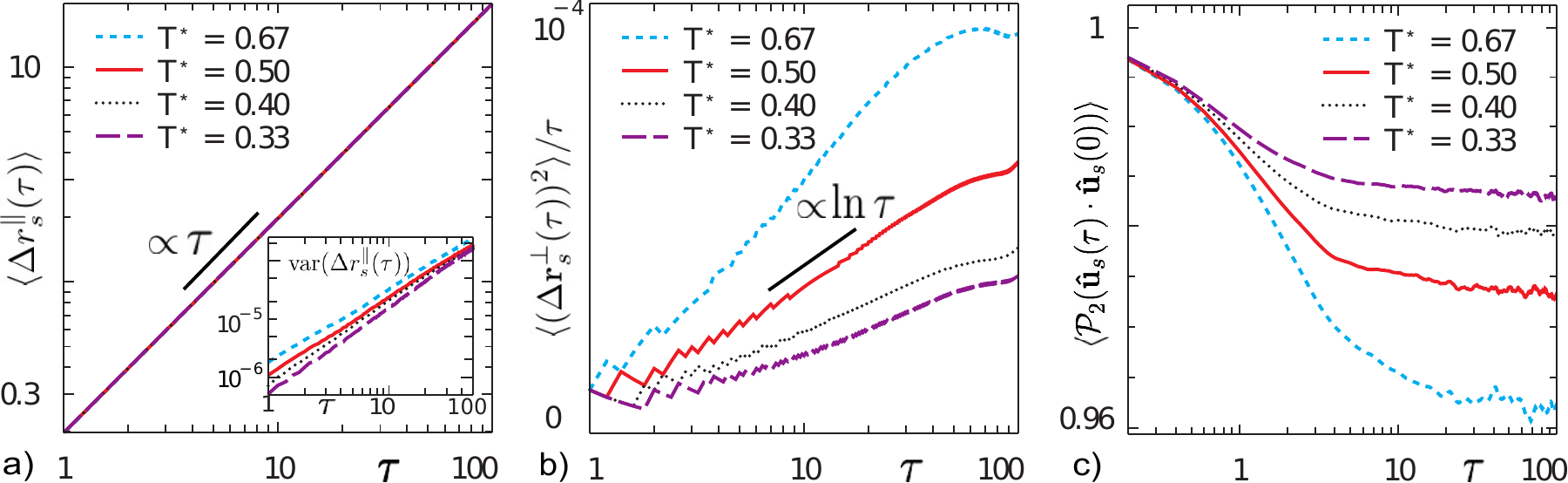}
\caption{Averaged displacements of an active particle moving in a nematic
medium characterized by a reduced temperature
$T^{\ast}$ (simulation parameters: $f_{0} = 5$, $N=50^{3}$, $ \sigma = 2a$
and $\ell =5$). Note the double-logarithmic scale. All distances are normalized in units of the box length. (a) Mean displacement along the nematic director  showing a linear, ballistic scaling.  Inset: variance in the parallel displacement  ${\rm var}(\Delta r_{s}^{\parallel}(\tau)) =  \langle ( \Delta \bfr_{s}^{\parallel}(\tau))^{2}  \rangle -\langle \Delta r_{s}^{\parallel}(\tau)  \rangle^{2}$ confirming  weakly off-ballistic corrections imparted by short-wavelength director fluctuations. (b) Mean squared displacement
transverse to the director  exhibiting anomalous logarithmic
scaling $\propto \tau \ln \tau$ (cf. \eq{theoll}). (c) ${\mathcal
P}_{2}$-weighted mean-squared rotation characterized by standard diffusive relaxation.}
\label{fig2}
\end{figure*}

By setting the lattice spacing $a$ as the internal length scale and defining $D_r = k_{B}T/\xi$ as the spin rotational diffusion coefficient  (not to be confused with the diffusion constant  for director reorientation $D\sim 3\varepsilon a^{2} / \xi $), 
we introduce dimensionless variables for time $\tau  = t D_{r} $, active force  ${f}_{0}  = \beta  F_{0} a$ 
and torque ${\bm \omega} = \beta {\bf \Omega} $. The reduced active force $f_{0}$ is also called P\'{e}clet number;
its order of magnitude follows from the typical thrust force of a microswimmer $F_{0} \sim 0.1 - 1$ $pN$ 
so that $|f_{0}| \sim 10^{1} - 10^{3}$. 
Therefore we arrive at reduced equations of motion:
\begin{align}
 d_{\tau} (\bfr_{s}/a)  &=  \tilde{\zeta}_{s}^{-1}   f_{0} \bhu_{s}  \nonumber \\ 
 d_{\tau} \bhu_{s} &= \tilde{\xi}_{s}^{-1}  (   {\bm \omega}_{s}   +  \hat{{\bm \omega}}_{s} )   \times \bhu_{s} 
\label{seom}
\end{align}
and a similar equation of motion for the Lebwohl-Lasher spins:
\beq
d_{\tau} \bhu_{i} =  ( {\bm \omega}_{i}    + {\bm \omega}_{i,s}  +    \hat{{\bm \omega}}_{i} ) \times \bhu_{i} 
\label{leom}
\eeq
We further introduced reduced friction coefficients $  \tilde{\zeta}_{s}  = \zeta_{s} a^{2}/ \xi  $ 
and $\tilde{\xi}_{s} = \xi_{s}/\xi$. If one assumes both for the swimmer and for the spins a spherical hydrodynamic shape 
(for the Stokes friction in a molecular solvent) one can approximate $ \tilde{\zeta}_{s}  \approx 4 \ell /3 $
and $  \tilde{\xi}_{s}   \approx  \ell ^{3} $
with $\ell$  denoting the ratio between the hydrodynamic radii of the active particle and the spin. 
The hydrodynamic size asymmetry $\ell$ sets the typical reorientation rate of the spin versus that of the swimmer.
Typically, $\ell >1$ for elongated swimmers.

The random torques correspond to Gaussian rotational fluctuations with relative strength 
$\sqrt{2} $ for the spins and  $\sqrt{2 \tilde{\xi}_{s}^{-1} } $ for the swimmer.  
 We defined the displacement vectors $\Delta  \bfr^{\parallel}_{s}(\tau)\equiv (\Delta \bfr_{s} (\tau) \cdot \bn) \bn $ along, and $\Delta  \bfr^{\perp}_{s}(\tau)\equiv  \Delta \bfr_{s} (\tau) - \Delta\bfr^{\parallel}_{s}(\tau)$ perpendicular to, the nematic director $\bn$,  with $\Delta \bfr_{s}(\tau) = \bfr_{s}(\tau)  - \bfr_{s}(0)$. We then determined from our simulations the
mean squared displacements  $\langle ( \Delta  \bfr^{\parallel}_{s}(\tau) )^{2} \rangle$ and $\langle ( \Delta  \bfr^{\perp}_{s}(\tau) )^{2} \rangle$, where $\langle \cdots \rangle$ denotes a time-average in the steady state.
Likewise we may define mean-squared rotation  via:
 \beq
 \langle   {\mathcal P}_{n} ( \bhu_{s}(\tau) \cdot \bhu_{s}(0)) \rangle, \hspace{0.5cm} n=1,2
\eeq
 in terms of Legendre polynomials ${\mathcal P}_{n}(x)$. 
 Recasting the theoretical prediction \eq{theo} into the units defined for the Lebwohl-Lasher model yields
\beq
 \langle ( \Delta \bfr_{s}^{\perp}(\tau))^{2} \rangle = \frac{1}{\pi} \frac{ f_{0} T^{\ast}}{ {\tilde{\zeta}_{s}}} \tilde{K}^{-1} \tau \ln \tau + {\mathcal O} (\tau)
 \label{theoll}
\eeq
We reiterate that  in the model the three Frank elastic constants are equal and reach the limiting value $ \tilde{K} = Ka/\varepsilon \rightarrow  3$ at zero temperature \cite{priest1972}. Finite temperature corrections have been quantified numerically in Ref. \cite{cleaver1991}. A reasonable fit of the simulation data is obtained using the following parameterization in terms of the first two nematic order parameters  
\beq
\tilde{K} \simeq c_{22} S_{2}^{2} + c_{24} S_{2} S_{4}
\eeq
with $S_{2} = S = \langle {\mathcal P}_{2} (\bhu \cdot \bn ) \rangle $ and $S_{4} = \langle {\mathcal P}_{4} (\bhu \cdot \bn ) \rangle $. The coefficients are $c_{22} = 3.905$ and $c_{24} = -0.905$. These parameters reproduce the exact result for the zero temperature case where both order parameters tend to unity.
 
 We have numerically solved the coupled equations of motion for the swimmer and the nematic medium by 
 a simple linearized scheme for \eq{seom} and \eq{leom} using a sufficiently small time-step $\delta \tau < 0.001$.
 We thereby generate trajectories for $\{  \bhu_{s}(\tau),  \bfr_{s}(\tau) ; \bhu_{i}(\tau)\} $ to perform the averages.
 An equilibration run of duration $\Delta \tau = 100$ starting from a system of perfectly aligned spins with a swimmer fixed at the centre of the system is followed by a  production run with a mobile swimmer during which statistics were gathered over a time interval of at least $\Delta \tau = 5000$.
 The system size is fixed at $N=50^{3}$ spins with periodic boundary condition in all three directions. 
 Spontaneous director rotation can be avoided by random spin flips at initiation to minimize the net 
spin magnetization  \cite{allen2005}.

The essential control parameters of our model are: i) the effective temperature $T^{\ast} = (\beta \varepsilon)^{-1}$ of the medium
which controls whether it is in an isotropic or nematic state and, in the latter case, the strength of the nematic director fluctuations.  ii) The
reduced active force (or P\'eclet number) $f_{0} >0 $. Provided sufficiently  large, this parameter is of minor importance for the scaling properties of the swimmer mobility. 
iii) The range  $\sigma $ over which the swimmer is influenced by  its nematic background and vice versa. iv) The ratio $\ell$ controlling the orientational relaxation of the background
 and the swimmer. For $\ell >1$ and $ \sigma >1$ the coupling between the swimmer and its nematic surrounding is strong enough to rule out any hairpin turns to occur within the explored simulation time.  Some typical examples of swimmer trajectories  generated from the simulations are depicted in Fig. 1.

\begin{figure*}
\includegraphics[width= 1.8 \columnwidth]{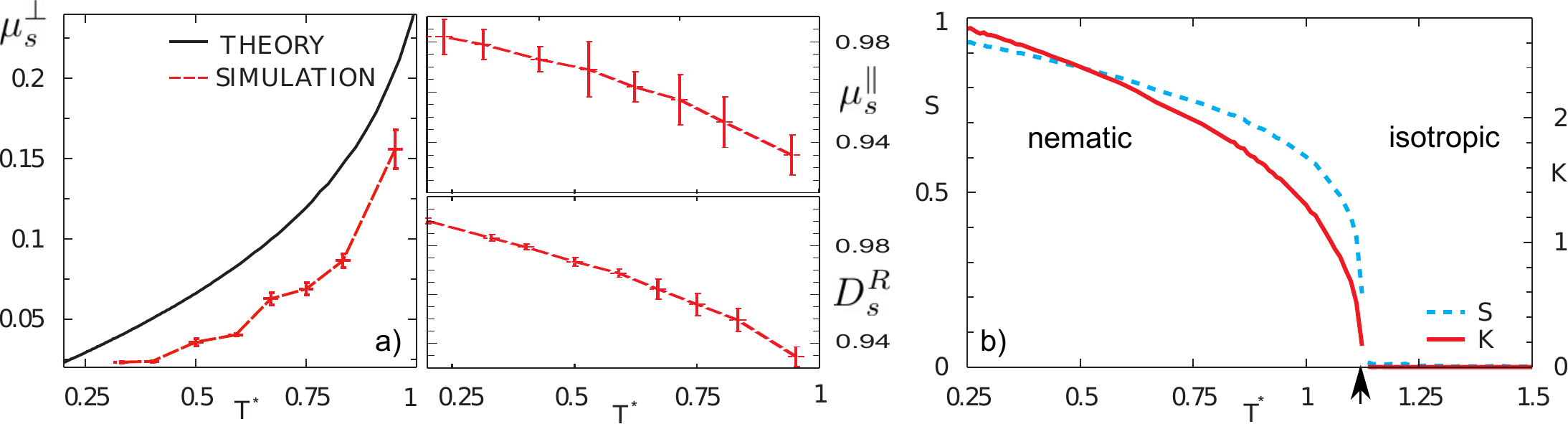}
\caption{Swimmer mobility versus temperature $T^{\ast}$ for the transverse
($\perp$), parallel ($\parallel$) and rotational (R) degrees of motion
with respect to the nematic director. The transverse and parallel
mobilities are defined as $\mu_{s}^{\perp} = \lim_{\tau \rightarrow \infty}
\langle (\Delta \bfr_{s}^{\perp} (\tau))^{2} \rangle /\tau \ln \tau$ and
$\mu_{s}^{\parallel} = \lim_{\tau \rightarrow \infty} \langle (\Delta
\bfr_{s}^{\parallel} (\tau))^{2} \rangle /\tau^{2} $, respectively. The
two translational contributions exhibit opposite trends with temperature
$T^{\ast}$. The rotational mobility is represented by the long-time
rotational diffusion constant of the swimmer,  defined as $ D_{s}^{R} = \lim_{\tau \rightarrow
\infty} -2 \ln \langle {\mathcal P}_{1} ( \bhu_{s}(\tau) \cdot
\bhu_{s}(0)) \rangle/ \tau $. (b) Bulk isotropic-nematic (IN) phase
diagram of the Lebwohl-Lasher  showing the nematic order parameter
$S$ and elastic constant $K$ (in units $k_{B}T/a$) versus temperature. A
weakly first-order IN transition at $T^{\ast}_{IN} \approx 1.12
$ is indicated by the arrow.}
\label{fig2}
\end{figure*}

 \section{Mean displacements and temperature-dependent mobility }

We start by testing the scaling predictions for the averaged displacements parallel and transverse to the nematic director for various temperatures in the nematic regime of the Lebwohl-Lasher spins  (the phase-diagram is shown in Fig. 3c).  While the parallel displacement is  ballistic  for all temperatures (Fig. 2a), the transverse contribution is highly non-trivial and displays the characteristic logarithmic long-time behavior borne out from our hydrodynamic theory (Fig. 2b). 
The rotational displacement functions depicted in Fig. 2c demonstrate that the swimmer quickly loses memory of its initial orientation but remains strongly aligned to the nematic director: this is enforced by the coupling term \eq{swimtorque}. Consequently,  ``hairpin" turns are absent  within the time frame of our simulations (see also Appendix A), and the swimmer keeps moving with its main direction parallel or anti-parallel to $\bn$, depending on its initial direction at $\tau=0$.

We now proceed with investigating the effect of temperature of the nematic on the swimmer mobility. The corresponding results are shown in Fig. 3a.  Reducing the temperature of the nematic increases the parallel mobility of the swimmer, however only slightly.  In the limit of zero temperature the average velocity is expected to yield the mean propulsion velocity $v_{s}$ of a free particle. The perpendicular mobility, however, {\it decreases}  with reducing $T^{\ast}$. This trend is captured qualitatively by the theoretical prediction \eq{theo}.
Finally, the long-time rotational motility, represented by the rotational diffusion constant of the swimmer $D_{s}^{R}$, decreases upon reduction of $T^{\ast}$ because the elastic forces dominate the thermal rotational fluctuations as $T^{\ast}$ drops.  

Our results demonstrate that,  in contrast to  isotropic media, the coupling between temperature and  the Frank elasticity of the nematic medium strongly influences the swimmer mobility, in particular the transverse component. This opens up new possibilities to control self-assembly and collective behavior of active particles and swimmers by fine-tuning their microscopic mobility through temperature.

\section{Conclusion}

Motivated by recent experimental studies of active agents moving complex and anisotropic media we have focused on analyzing the basic problem of diffusion of an active particle or microswimmer in a macroscopic nematic liquid crystal.  
Using hydrodynamic scaling theory complemented by numerical simulation 
we have explored self-diffusion of a self-propelled particle whose motion is affected 
by the thermal orientational fluctuations of the nematic background. These fluctuations couple to the swimming direction and therefore
induce a superdiffusive motion perpendicular to the nematic director field. The amplitude of this motion can be varied by changing the thermodynamic
parameters characterizing the background, in particular the temperature of the nematic. 

Our predictions
 can be exploited to control and tune the anisotropic motion of active carriers 
in non-Newtonian fluids which is of importance e.g. 
in drug delivery. Our results should be verifiable 
in real-space experiments on swimmers, both artificial ones or bacteria. The liquid crystalline medium can either 
be a molecular liquid
crystal \cite{PNAS_Aranson,Sagues,PNAS_Aranson} or a passive colloidal liquid crystal.

Future work should be aimed at generalizing various aspects of the model used in this work. First of all, it would be intriguing to address the behavior 
of a collection of microswimmers moving in a nematic medium. While for isotropic, Newtonian solvents, collective effects such as swarming, kinetic clustering and turbulence
\cite{Marcetti_RMP, romanczuk2012, Gompper_Winkler_review, buttinoni2013} have been firmly established, it is unkown how this behavior is altered or enriched in case of collective propulsion in a nematic liquid crystalline medium.

Moreover the two-dimensional
case (as e.g. realized for a passive colloidal monolayer hosting a swimmer on a substrate)
can be studied, for which the thermal fluctuations of the director field are much larger than in three 
spatial dimensions \cite{MerminWagner,Vink2009}. We have analyzed this problem by the techniques  used here,  and find isotropic super diffusive behavior:
\beq
 \langle | \Delta \bfr_{s}(t)|^{2} \rangle \propto t^{\Upsilon(T)} \quad ,
 \label{theo2d}
\eeq
where the non-universal, temperature-dependent exponent $\Upsilon(T)$ is given by 
\beq
\Upsilon(T) ={4\over 2+\eta(T)} \quad ,
 \label{upsilon}
\eeq
where $\eta(T)$ is the non-universal, temperature dependent exponent characterizing the algebraic decay of director correlations in the low temperature, ``Kosterlitz-Thouless" \cite{kosterlitz1973} phase of the 2D nematic, which exists for temperatures $T<T_{KT}$, where $T_{KT}$ is the Kosterlitz-Thouless transition temperature, above which director fluctuations become short-ranged (and the swimmer motion becomes conventionally diffusive). The exponent $\eta(T)$ is usually a monotonically decreasing function of temperature, and is always \cite{jkkn1977} bounded: $0\le\eta(T)\le{1\over 16}$. Thus, the exponent $\Upsilon(T)$ will usually be  a monotonically decreasing function of temperature, and will always have a very narrow range of variation:  ${64\over 33}=1.939393...\le\Upsilon(T)\le 2$. Details of this calculation will be given in a future publication \cite{future}.

\section*{Acknowledgments}

Helpful discussions with F. Sagues, P. Tierno, I. Aronson and J. Lintuvuori are gratefully acknowledged. This work originated from the KITP program, ``Active Matter: Cytoskeleton, Cells, Tissues and Flocks". HL acknowledges funding from the German Research Foundation (DFG) within the Priority Program  {\it ``Microswimmers"} (SPP 1726). J.T. also thanks the Max Planck Institute for the Physics of Complex Systems (MPI-PKS), Dresden, Germany;  and  Chiu Fan Lee and the Department
of Bioengineering, Imperial College, London, UK for their hospitality while this work was underway.  He also thanks the  US NSF for support by
awards EF-1137815 and 100617. \\

\section*{Appendix: Hairpin Turns} 

In this Appendix, we consider hairpin turns. In Part A, we estimate the characteristic time  $t_{\text{hairpin}}$ between such turns, and show that, at the temperatures of our simulations, it should be astronomically  large (making these events far less frequent than black hole mergers!). This is consistent with our observation that such turns {\it never} occur in our simulations. In Part B, we show that for time scales $t  \gg t_{\text{hairpin}}$, both the mean drift along the nematic director, and the anomalous $t\ln t$ lateral diffusion, are lost, and motion in both directions becomes diffusive, albeit with different diffusion constants.

\subsection*{A. Rate of hairpin turns}

We begin by noting that the {\it rotational} dynamics of the swimmer are essentially equilibrium, with Hamiltonian
\beq
H = -\varepsilon_s  \sum_{j}  {\mathcal P}_{2} ( \bu_{s} \cdot \bu_{j} )g({\bf R}_j-\bfr_s) \,.
\label{LLswim}
\eeq
Since the difference between the minimum and maximum values of  $\mathcal P_{2} (x)$ (which occur at $x=1$ and $x=0$ respectively, corresponding to angles of zero and 90 degrees between the swimmer and the spin) is $3/2$, the minimum energy barrier against a hairpin turn is given by
\beq
E_B^{min}={3\varepsilon_s\over2}\left[\sum_{j}g({\bf R}_j-\bfr_s)\right]_{\rm{min}(\bfr_s)} \quad ,
\label{EB1}
\eeq
where $\left[f(\bfr_s)\right]_{\rm{min}(\bfr_s)}$ denotes the minimum value of any function $f(\bfr_s)$ over all $\bfr_s$.
On symmetry grounds, for the function $\sum_{j}g({\bf R}_j-\bfr_s)$, this minimum occurs when $\bfr_s$ is at the center of a unit cell. Hence, the Gaussian form  $g(r) = \exp[-(r / \sigma)^{2}]$, combined with writing the set of lattice positions ${\bf R}_j$ with the usual lattice indexing  ${\bf R}_j=ma\bxh+na\byh+pa\bzh$, with $m$, $n$, and $p$ running over all integers from $-\infty$ to $\infty$, and taking $\bfr_s=(a\bxh+a\byh+a\bzh)/2$ implies

\begin{widetext}
\begin{eqnarray}
\left[\sum_{j}g({\bf R}_j-\bfr_s)\right]_{\rm{min}(\bfr_s)}&=& \sum^\infty_{m=-\infty}\sum^\infty_{n=-\infty}\sum^\infty_{p=-\infty}  \exp\left(-[(m-1/2)^2+(n-1/2)^2+(p-1/2)^2]\left({a\over \sigma}\right)^2\right)\nonumber\\
&=&\left( \sum^\infty_{m=-\infty}  \exp\left(-(m-1/2)^2\left({a\over \sigma}\right)^2\right)\right)^3 \quad ,
\label{gmin2}
\end{eqnarray}
\end{widetext}
where to obtain the second equality we have used the associative property of multiplication and addition.

The sum 
\beq
h(k)\equiv\sum^\infty_{m=-\infty}  \exp\left(-k(m-1/2)^2\right)
\label{hdef}
\eeq
with $k\equiv\left(a / \sigma\right)^2$ in this last expression can be evaluated using the Poisson summation formula
\begin{eqnarray}
\sum^\infty_{m=-\infty}  f(m)=\sum^\infty_{s=-\infty}  \int_{-\infty}^\infty f(x)e^{2\pi i sx} dx\quad .
\label{fishy}
\end{eqnarray}
Applying this to the function $f(x)\equiv \exp\left(-k(x-1/2)^2\right)$ gives
\begin{eqnarray}
h(k)=\sum^\infty_{s=-\infty}  \int_{-\infty}^\infty \exp\left(-k(x-1/2)^2\right)e^{2\pi i sx} dx \quad .
\label{fishy2}
\end{eqnarray}
The Gaussian integral in this expression is easily evaluated, yielding
\begin{eqnarray}
h(k)=\sqrt{\pi\over k}
\left(\sum^\infty_{s=-\infty}  (-1)^s\exp\left(-{\pi^2s^2\over k}\right)\right)\quad .
\label{hsmallk}
\end{eqnarray}
This sum on $s$ in this expression converges extremely rapidly for any small $k$; indeed, simply keeping the leading order $s=0$ term is accurate to a part in $10^4$ for any $k>1$. In our simulations, we choose $\sigma=2a$, so $k=1/4$, for which keeping the leading order term is accurate to a part in $10^{17}$. We will therefore keep only the leading order term, which amounts to taking
\begin{eqnarray}
h(k)=\sqrt{\pi\over k}=\left({\sigma\over a}\right)\sqrt{\pi}
\quad .
\label{hsmallk}
\end{eqnarray}
Using this in (\ref{gmin2}) gives
\begin{eqnarray}
\left[\sum_{j}g({\bf R}_j-\bfr_s)\right]_{\rm{min}(\bfr_s)}&=& \left({\sigma\over a}\right)^3\pi^{3/2} \quad ,
\label{gmin2}
\end{eqnarray}
which can be used in (\ref{EB1}) to obtain our final expression for the minimum energy barrier against a hairpin turn:
\beq
E_B^{min}={3\varepsilon_s\over2}\left({\sigma\over a}\right)^3\pi^{3/2}  \quad .
\label{EB2}
\eeq
We emphasize that this expression only applies for $\sigma\gtrsim a$.

Evaluating it for the value $\sigma=2a$ used in our simulations gives 
\beq
E_B^{min}=12\varepsilon_s\pi^{3/2}  \quad .
\label{EBsim}
\eeq
Now since the rotational dynamics is essentially equilibrium, we expect the time for a hairpin turn to be of order the Boltzmann factor associated with the energy barrier, times the microscopic rotation time of the swimmer   $ \xi_s/ \kbt$.   That is:
\begin{eqnarray}
t_{\text{hairpin}}\sim \frac{\xi_s}{\kbt} \exp\left({E_B\over\kbt}\right)> \frac{\xi_s}{\kbt}\exp\left({E_B^{min}\over\kbt}\right) 
\label{th1}
\end{eqnarray}
Converting to the time units used in our simulation, and recalling the definition of  reduced temperature  $T^*=\kbt/\varepsilon_s$ we obtain
\begin{eqnarray}
\tau_{\text{hairpin}} > \ell^3 \exp\left({3\pi^{3/2}\over 2T^*}\left({\sigma\over a}\right)^3\right) 
\label{th1}
\end{eqnarray}

Thus, for the simulation we performed with the highest $T^*$, namely $T^*=0.67$ and, hence, the smallest value of $\tau_{\text{hairpin}}$, we obtain, using our simulation values $\sigma=2a$ and $\ell=5$, a lower bound on $\tau_{\text{hairpin}}$ of $\tau_{\text{hairpin}}>\ell^3\exp(18\pi^{3/2})=4.23\times 10^{45}$. For our simulations at lower temperatures $T^*$, the time between hairpins is even longer. Thus, it is hardly surprising that we see no hairpin turns in our simulations. Indeed, for a wide range of realistic values of the parameters, they will simply not occur on any time scale  accessible either in simulations, or in  experiments. For such parameters, therefore, the theory presented in the main text, which ignores hairpin turns, will be valid. In particular, both the ballistic motion along ${\bf \hat{n}}$ and  the logarithmically anomalous superdiffusion  transverse to ${\bf \hat{n}}$ should occur.

\subsection*{B.  Effect of hairpin turns}

Due to the exponential sensitivity of the hairpin turn time to various microscopic model parameters,  hairpin turns should occur in some experimental situations (or simulations) on a reasonable time scale. It therefore behooves us to consider their effect. We will argue in this subsection that hairpin turns,  for time scales $t  \gg t_{\text{hairpin}}$, destroy  both the mean drift along the nematic director, and the anomalous $t\ln t$ lateral diffusion. Instead, on these long time scales, motion in both directions becomes diffusive, albeit with different diffusion constants.

Our argument begins by modifying equation (\ref{align}) to include the possibility of hairpin turns:
\beq
{d \bfr_{s}(t)\over dt}=\Upsilon(t) v_s\bn(\bfr_{s}(t),t)+\bff(t) \,,
\label{alignflip}
\eeq
where
\beq
\Upsilon(t) =\pm 1 \,,
\label{upsdef}
\eeq
is a fluctuating Ising variable that changes sign every time a hairpin turn occurs. If we assume, as seems reasonable for a  with no long-term memory, that flips in the sign of $\Upsilon(t)$ are a Poisson process with rate $1/t_{\text{hairpin}}$, then we expect correlations of $\Upsilon$ to decay on a time scale $t_{\text{hairpin}}$:
\beq
\left<\Upsilon(t^\prime)\Upsilon(t^{\prime\prime})\right>=\exp(-|t^{\prime}- t^{\prime\prime}|/t_{\text{hairpin}}) \, .
\label{Upscorr}
\eeq
Thus we see that, even though the director $\bn(\bfr_s(t), t)$ has long-ranged temporal correlations, as discussed in the main text, correlations of the product $\Upsilon(t) v_s\bn(\bfr_{s}(t),t)$ will decay rapidly (exponentially) for $t\gg t_{\text{hairpin}}$. Thus, the anomalous behavior arising from the alignment of the swimmer velocity with the nematic director, namely, the development of a non-zero mean velocity along $\bn$, and the anomalous logarithmic superdiffusion ($t\ln t$). Instead, we expect the motion along the mean director direction will now consist, roughly, of a ``drunkards walk" along the mean nematic director $\langle\bn\rangle$ in which each step has mean length $v_z t_{\text{hairpin}}$, and lasts a mean time $t_{\text{hairpin}}$. This is readily seen to lead, on longer time scales ($t\gg t_{\text{hairpin}}$), to diffusive behavior with a diffusion constant 
\beq
D_{s}^{\parallel}=v_z^2t_{\text{hairpin}} \,.
\label{Dpar}
\eeq
The transverse wandering $ \langle ( \Delta \bfr_{s}^{\perp}(t))^{2} \rangle$ will have its $\ln t$ factor, which arises from the long time correlations, cut off by $t_{\text{hairpin}}$ for $t\gg t_{\text{hairpin}}$, leading to diffusive behavior in that direction as well, but with a diffusion constant
\beq
D_{s}^{\perp}\sim\frac{v_{s} k_{B}T}{2\pi} (K_{1}^{-1}+K_{2}^{-1}) \ln \left(\frac{t_{\text{hairpin}}}{  t_{0}}\right)\,.
\label{Dperp}
\eeq
Note that $D_{s}^{\parallel}\gg D_{s}^{\perp}$ for very large $t_{\text{hairpin}}$.

\bibliographystyle{apsrev4-1}
\bibliography{refs}

\end{document}